\documentclass[11pt]{article}
\usepackage{graphicx}
\usepackage{amsmath}
\usepackage{amssymb}
\usepackage{amsxtra}
\usepackage{changebar}
\usepackage{placeins}
\newcommand{\bc}{\begin{center}}
\newcommand{\ec}{\end{center}}
\newcommand{\bd}{\begin{displaymath}}
\newcommand{\ed}{\end{displaymath}}
\newcommand{\be}{\begin{equation}}
\newcommand{\ee}{\end{equation}}
\newcommand{\ba}{\begin{array}}
\newcommand{\ea}{\end{array}}
\newcommand{\bea}{\begin{eqnarray}}
\newcommand{\eea}{\end{eqnarray}}
\newcommand{\bt}{\begin{tabular}}
\newcommand{\et}{\end{tabular}}

\newcommand{\bp}{\begin{picture}}
\newcommand{\ep}{\end{picture}}
\newcommand{\bfi}{\begin{figure}}
\newcommand{\efi}{\end{figure}}
\begin{document}

\title{{\huge \bf Simple Mass-estimates 
for Resonance(s) being 
 6 Top plus 6 Anti 
top Bound states and Combinations 
thereof}}

\author{
H.B.~Nielsen
\footnote{\large\, hbech@nbi.dk, 
hbechnbi@gmail.com}
\\[5mm]
\itshape{
The Niels Bohr Institute, Copenhagen,
Denmark}\\
}

\date{}

\maketitle

\begin{abstract}
We have long speculated
\cite{1nbs,2nbs,3nbs,4nbs,5nbs,
6nbs,7nbs,8nbs,9nbs,10nbs,11nbs,12nbs,
13nbs,14nbs,LNvacuumstability}
, that 
6 top + 6 anti-top quarks due to the 
realtively large size of the top-yukawa 
coupling would bind exceptionaly 
strongly by mainly Higgs exchange.  
Here we 
present a surprisingly simple 
``calculation'' of the mass of 
this speculated bound state. 
Even a possible resonance in scattering 
of two such bound states is speculated. 
For the ``calculation'' of the masses 
it is crucial to assume, that our
since long speculated principle ``Multiple
Point Principle''\cite{5mp,6mp,7mp,8mp,9mp,10mp,
11mp,12mp,13mp,14mp,15mp,16mp,17mp,
deriving},  is true.
This principle  says: {\em there are 
several vacua all having almost zero 
energy density}. Further we make an 
approximation of the Higgs Yukawa 
potential essentially replacing the 
exponential in it by a step-function.
The new result means that there  
are now  two independent calls 
for our bound state having the mass 
around  750 GeV required by  our 
``new law 
of nature'' the Multiple Point Principle.
It should be remarked that in our 
picture there is {\em no new physics} in 
the sense of 
new fundamental particles, but the 
``Multiple Point Principle'' is new in the 
sense of being not yet accepted. 
Further we get the {\em same} mass within 
uncertainties as earlier\cite{LNvacuumstability} but now from 
a completly different
assumption, except for  being 
from 
our ``multiple point principle''. But 
the two masses are gotten from using 
 {\em different}  (speculative) 
vacua occuring in the pure Standard 
Model. 
\end{abstract}

\section{Introduction}
We - especially C.D. Froggatt 
Larisa Laperashvili and myself -
have long been speculating on a 
very strongly bound state \cite{1nbs,2nbs,
3nbs,4nbs,5nbs,
6nbs,7nbs,8nbs,9nbs,10nbs,11nbs,12nbs,
13nbs,14nbs,LNvacuumstability, FN750} of 
6 top and 6 anti top quarks, 
being held together mainly by 
exchange of Higgs bosons in a 
picture connecting it with our 
(and Don Bennetts also) 
principle of degenerate vacua
\cite{5mp,6mp,7mp,8mp,9mp,10mp,
11mp,12mp,13mp,14mp,15mp,16mp,17mp,deriving}. Mostly we just claimed that we 
assumed/imagined, that this bound
state were very strongly bound, 
meaning that its mass should be 
appreciably smaller than the 
collected mass of its 12 
constituents $12m_t = 2076 GeV$.
However, C. D. Froggatt and 
I\cite{Tunguska} in some 
appendices mainly achieved an 
estimate of the order of 285 GeV
for the mass of this bound state.
This estimate were strongly based
on the assumption of the 
degeneracy of the present vacuum 
with a vacuum, in which there is 
a lot of copies of the bound 
state, the ``condensate'' vacuum.
It is the main purpose of the 
present article to redo in a 
somewhat more direct way the 
calculation, made together with 
Colin Froggatt. As a little fun 
thing some corrections 
approximately cancel each other 
in such a way that we can get a 
rather simple calculation at the 
end, and the very simple result 
that the mass of the bound state
very crudely has the mass $4m_t$,
where $m_t$ is the mass of the 
top quark.   

The great interest of this 
crude estimation of the mass for 
our hoped for bound state from 
the mentioned degeneracy 
assumption of the ``present''
and the ``condensate vacuum''
is that recently Laperashvili,
 Das, and myself connected
\cite{LNvacuumstability, FN750} 
the 
mass of our bound state with the 
rather little difference in 
energy density between the 
present vacuum, and the vacuum 
corresponding to the high Higgs 
field minimum in the effective 
potential for the Higgs field.
And remarkably, we found within 
estimation uncertainty the same 
mass in the two different ways 
of estimating the mass of the 
bound state. It should be 
stressed and understood, that we 
have a picture, in which we 
speculate there to exist three 
- and most importantly degenerate
- vacua, and then we calculate 
the mass of our bound state in 
two different ways, namely 
requiring the present vacuum 
degenerate with the 
``condensate'' giving - in the 
present article - the value 
4$m_t$, and with the ``high Higgs
field vacuum'' giving about 
710 GeV or 850 GeV also rather 
uncertain though. It is the 
remarkable compatibilty of these
two quite independent mass 
estimates, which is our main 
point! If this is not just 
an accident, then there must be something about both the bound 
state and our Multiple Point 
Principle about degenerate vacua! 
 
This ``Multiple Point Principle'' 
causes restrictions 
between coupling constants and 
thus 
potentially serves as a 
candidate for a 
``solution'' to  fine tuning 
problems\cite{6nbs,
  7nbs}. This ``Multiple Point Principle''
(=MPP) says, indeed,
that there are several vacua with 
extremely small energy densities.
We could also say, that it means 
that 
the universe-vacuum is just at 
some 
multiple point, where several 
phases 
can coexist, much like one at 
the triple 
point for water has coexistence 
of 
ice, fluid water, and vapor for a common 
set of intensive variables, 
pressure and 
temperature. There may be no real 
good derivation or argument for 
our 
multiple point principle in 
spite of 
the fact, that we and others have 
publiched some 
attempts to derive this principle
\cite{5mp,deriving, Stillits, 
Kawana1, Kawana2},
but all such arguments would
have to involve some influence 
of the 
future on the passed,or  at least on the 
coupling constants, and that 
would make all
such derivations of MPP 
(=``Multiple Point Principle'') 
suspicious.
The reader should rather take 
some 
previous works - even 
{\em pre}diction(s) - 
as well as the results of the 
present 
work as {\em empirical} 
evindence for 
this new law of nature, 
the`` multiple 
point principle''. 
  
The calculation to be presented below 
is indeed just a slightly renewed version 
of the calculation delivered in the 
appendices of our earlier 
article \cite{Tunguska}, in which it is 
heavily 
used, that there should exist a new vacuum 
degenerate in energy density with the 
present one and with an approximated 
structure, as if the S-particles(what we 
called our bound state, 
which so much happened to fit the 
by now digamma fluctuation 
called in its fashion F, that we 
shall call it F now 
) were 
sitting as the carbon atoms in the diamond 
chrystal, as we shall review in section 
\ref{modelling}. Colin Froggatt and 
I ended in these appendices\cite{Tunguska}
 with a mass 260 GeV for the bound state,
but we certainly did not believe our mass 
{\em pre}diction 260 GeV to be 
very accurate. For the trustworthiness 
of our whole story the recent work of 
Larisa Laperashvili, Chitta Das and myself
\cite{LNvacuumstability} relating the 
{\em mass of} the speculated {\em bound 
state to} 
the degree of instability/{\em negative 
energy density} of the second minimum in 
the Standard Model Higgs field effective 
potential calculted without inclusion 
of our bound state. The point is that 
in order to achieve just zero energy 
density (as our Multiple Point Principle 
requires) for the vacuum represented by 
the 
second minimum the correction required
is just getting right for the mass of the 
bound state F 
being in a range very 
close to 750 GeV. We shall return to 
this work in the last subsection in the 
conclusion \ref{Larisaetal}.

In the following section \ref{boundMPP}
we shall review our model of there 
existing an exceedingly strongly bound 
system of 6 top + 6 anti top quarks, 
and of our ``multiple point principle'' 
fine tuning the coupling constants, so 
that for instance a condensate of the 
bound state can fill the vacuum and 
cause a ``new vacuum'' with the 
energy density just finetuned to be again 
remarkably small, of the same order as 
say the astronomical observation of the 
energy density(= cosmological constant)
of the vacuum, in which we live.
(This astronomically observed cosmological
constant is quite negligible compared 
to the energy densities of any 
significance for high energy physics 
parameters such the bound state mass or
the Higgs mass). 
A subsection \ref{MPP} of this section 
\ref{boundMPP} is assigned to 
our ``new law of nature'',``Multiple 
point
principle''. In the next 
section \ref{modelling}
we then model in a very crude 
approximation 
and in a non-relativistic picture 
the just mentioned new vacuum\footnote{ 
Our full 
picture has actually even more vacua, e.g.
one more in pure Standard Model here 
called 
``High Higgs field vacuum'', because 
the Standard Model Higgs in that vacuum 
has a magnitude of the order of $10^{18}\;  GeV$} in our model called the 
``condensate vacuum'' as containing a
bose-condensate of the F  bound states  
by suggesting 
as a very crude approximation, that this
vacuum has a system/a lattice of F(750)
particles interacting  
with their neighbors contained in the 
vacuum. We take the ``atoms''= the F's 
in  this lattice to  
interact in analogy with 
the carbon atoms in a diamond chrystal.
In order to compare the interactions 
and the binding energies we ignore the 
effect, that when a top quark  goes around 
/is bound to a swarm of with the same 
radius bound quarks and antiquarks, it 
only ``feels'' the force from about half 
the number of particles in the swarm.
However, we argue in section 
\ref{accident}, that the exchange of 
what we call ``eaten Higgses'', and which  really is  exchange of the longitudinal 
components of  weak gauge 
bosons W and Z, {\em happens}  
with help of gluons also just 
accidentally to  
cancel this effect. It is very 
important for the success of our whole 
picuture of bound states and a vacuum 
condensate numerically, that inside the 
bound state F as well as in the 
condensate vacuum the effective Higgs-mass
is appreciably lower than the Higgs mass
of 125 GeV observed experimentally. 
Since, however, the Higgs field 
expectation value 
inside the bound 
state and inside the condensate vacuum     is significantly lower than in the usual 
vacuum,
the Higgs self-interaction indeed cause 
a smaller Higgs mass effectively in these 
places 
with many top and anti top around  
on the average. This deminished effective
Higgs mass is discussed  in 
section \ref{effHiggsmass}.
In section  \ref{potentialapproximation} 
we then 
for simplicity make the very crude 
assumption of approximating the 
exponential factor 
$\exp{(- m_{\hbox{effective Higgs mass}}r)}$ 
in the Higgs-Yukawa-potential by a 
step-function, a $\theta (``number'' - 
 m_{\hbox{effective Higgs mass}}r)
$ meaning, that we put the 
Higgs {\em mass} to zero for small 
distances, 
while we put the Higgs {\em Yukawa 
potential} 
to zero for large distances.
Next in section \ref{mass750} we 
``calculate'' or rather very crudely 
estimate the mass of the bound state 
F {\em from the 
requirement of the MPP assumption of 
the equality of the energy density 
of the ``condensate vacuum'' and the 
vacuum, we live in}. So our new principle 
MPP is really crucial for our mass 
prediction!

In section \ref{mass1c8} we also with the 
same picture discuss the at LHC actually 
first 
found possible resonance -of  mass 
1.8 TeV - seen(?) decaying into weak gauge 
bosons(it is very dobtful). We  take 
this resonance to be 
composite from a couple of F's as 
very weakly suggested from the mass of the 
1.8 TeV (we shall see combining 
the present work with our earlier
work with Laperashvili and Das 
\cite{LNvacuumstability} that an 
F-mass around 800 GeV is called for)
 possible particle being crudely 
twice that of F.  A little problem
for our interpretation of the 1.8 TeV 
state this way may be its relatively 
small width observed. The problem is
 dicussed 
a bit in the subsection \ref{Width}.

In section \ref{conclusion} we review 
and comment our result. In the 
subsection \ref{Larisaetal} we 
summarize, that the value for the bound 
state estimated in 
the present article - developping the 
result of \cite{Tunguska} - and the value 
for the bound state needed for a quite 
different multiple point principle 
requirement coincides remarkably!
       
\section{Bound State Picture and 
``Multiple 
Point Principle''}
\label{boundMPP}
The crucial suggestion behind our bound 
state model of 6 tops + 6 anti tops 
that since Higgs 
exchange 
like any other even order tensor particle 
exchange delivers attracktion between 
top and top, or top and anti top, or 
anti top and anti top as well, we 
get stronger and stronger binding between 
the 
top and anti top quarks the more of them 
we imagine brought
together. It is because the top and anti 
top are the strongest binding quarks, that
this type of binding becomes most relevant
for the top and anti top. Now. however,
the quarks are fermions and thus you 
cannot just unlimmited clump
arbitrarily many, e.g. top quarks, 
together.
Since the top quark has a color degree
of freedom taking three values, say: red, 
blue,
and yellow, and a spin degree of freedom,
that can be up and down, one can bring 
{\em 3*2=6 top quarks into the  same 
orbital
state}, but because of fermi-statistics 
no more. So there can in a single orbital
state be just up to 6 top + 6 anti top.
Thereby a {\em closed shell} is so to 
speak 
formed (in the nuclear physics sense).
In the zero Higgs mass approximation,
which will be effectively valid, when the 
size of the bound state - the 
radius - multiplied by the effective 
Higgs mass is small, the attracktion 
between the top-quarks or between 
tops and anti tops is quite analogous  
to that between an atomic nucleus and an
electron. So  we can for first 
orientation use the terminology from 
the quantum mechanical description 
of atomic physics. Approximating the 
bound state, that we suggest to be 
possible to form from  6 top + 6 anti top 
by thinking
of each top or anti top going arround a 
collected object formed from the other 
11 quarks, we can talk about different 
``orbits'' in the atomic terminology 
of a main quantum number $n$ taking 
positive integer values and further $l$
(the orbital angular momentum magnitude 
being $\sqrt{l(l+1)}$) 
and $m$(the angular momentum around the 
quantization axis). As in atomic physics 
the 
particles in the n=1 orbit are 
bound strongest. Analogous to the 
helium atom having especially high 
excitation energies, we have because 
of the color factor 3 and both quark 
and anti quark an especially stable 
system being a bound state of 6 top 
and 6 anti top quarks. 

Whether the binding of such a system
of 6 top + 6 anti top now is sufficiently 
strong to even bind 
to form a resonance, 
with 
a 
rather small mass (as we shall need say 
about 750 GeV) compared to 
the collective mass of 6 top + 6 anti top,
 12 $m_t$ = 12* 173 GeV = 2076 GeV, is 
controversial\cite{Kuch1,Kuch2,Kuch3}.
However, we think 
ourselves\cite{4nbs},
that making use of a long series of 
corrections, especially also exchange 
of the other three components of the 
Higgs than the as simple particle 
observed component, we can stretch 
the uncertanties in the calculation 
so far as to allow such a light bound 
state to be possibly 
formed\cite{4nbs}. 
These other components
of the Higgs are really present in the 
Standard model as W's and Z longitudinal 
components. We call them ``eaten Higgses'',
but really of course it just means to 
include weak gauge particle Z and W  
exchanges.
 
It is important for our hope, that the 
bound state can indeed bind so strongly, 
that it gets so tightly bound, that the 
strong Higgs fields inside the bound state
even can modify the effective mass of the 
Higgs significantly there. {\em We} 
estimated that 
a top-Yukawa coupling of $g_t 
=1.02 \pm 14 \%$ would be just sufficient
to bind an extremely light bound state 
of the 6 top and 6 anti top, and would 
match with the experimental top-Yukawa
$g_t =0.935$. But Shuryac et al. 
\cite{Kuch1, Kuch2,Kuch3}
find, that due to the high Higgs mass, it 
cannot bind at all for the experimental 
value of $g_t$.
     
\subsection{MPP}
\label{MPP}
The whole speculation about our bound 
state of 6 top + 6 anti top is a priori 
rather much taken out of the air by itself.
However, we have all the time proposed 
it only connected with another speculation,
the ``Multiple Point Principle''. This 
is, you
could say, a wild guess about simplifying 
the fine tuning problems of the 
Standard Model. In order to formulate 
just the cosmological constant problem
about, why the cosmological constant(= 
the vacuum energy density) compared to 
say Planck scale dimensional expectations
is so enormously small, one needs an 
{\em assumption} of the form ``The energy
density of vacuum is extremely small!''
Now you could look at the ``Multiple Point 
Principle'' as an extension of this anyway
needed assumption, without really 
complicating it severely: ``Several 
vacua have extremely small energy 
densities!''\footnote{I thank 
Leonard Susskind for the 
 remark behind this argument for 
MPP}. We almost just have 
put the anyway needed assumption into 
``plural'', or
changed the ``quantor'' from  
``The physical vacuum...'' to ``Several
vacua...''.

Now the real supporting point for this 
principle is, that although it is not 
unneccesarily 
complicated, it  is the one, which Colin 
Froggatt and 
myself managed to use to make 
historically\cite{9mp} in 1996, long 
before the Higgs 
particle were found, to make  a 
{\em pre}diction of
the Higgs mass of 135 GeV $\pm$ 10 GeV.
Now our prediction using the same
Multiple Point Principle would be 
129.4 GeV\cite{Deg} but with a much 
smaller 
uncertainty, comparable to the 
experimental 
uncertainty of a few hundred MeV. So 
at first it then looks, that while our 
original {\em pre}diction agreed perfectly
within errors, and 
the Multiple Point Principle were 
perfectly right, it is today 
deviating
of the order of three standard deviations
from matching experiment.
This formal disagreement of the 
theoretical prediction 
actually occured in spite of, that the 
better 
calculations and better
top mass moved our prediction {\em closer}
 to 
the experimental
value 125 GeV during the time we had 
predited it. It is of  course only 
possible that in spite of this development the agreement relative to the uncertainty 
could become worse, because the 
uncertainties in 
calculation and top and Higgs masses 
went down even faster. However, L.V. 
Laperashvili, 
C. Das and myself\cite{LNvacuumstability}
 found, that 
the existence of the bound state F 
of the 6 top + 6 anti top would make a 
{\em little theoretical correction to 
the mass 
of the Higgs}
being predicted from the multiple point,
so that the agreement might indeed 
be improved to be perfect, if the 
mass of this bound state is appropriate. 
A mass of the bound state $\sim 800$ would 
be fine for correcting the Higgs mass to 
be observed to agree with a perfect 
degeneracy of the vacua.

It should be stressed, that this 
successfull Higgs-mass {\em pre}diction 
as well as Colin D. Froggatts and mine 
controversial argument, that the 
top-Yukawa-coupling $g_t$ in order to 
allow for a condensate of bound states of 
6 top and 6  antitop with energy density 
close to zero, must be close to the 
value $1.02 \pm 14 \%$ 
supports the ``Multiple Point 
Principle'' as being a principle 
uphold by nature. The value 
$g_t = 1.02 \pm 14 \% $ namely matches
with the experimentally determined 
Higgs Yukawa coupling $g_t = 0.935$. 
Really we just estimated, what 
the top-Yukawa coupling should be in order,
that the bound state assumed to exist 
of 6 top + 6 anti top should have 
exceptionally low mass. But this should be 
approximately 
needed to have the condensating particle 
have mass close to zero in order for there
being two degenerate vacua as required 
by MPP(=``Multiple Point Principle'').

This means, that even if the theoretical 
arguments for the MPP are not totally 
convincing, then there is some {\em 
empirical 
evidence} pointing in favor of this MPP.    And the present article is meant to 
provide one more such indirect  
phenomenological support for MPP.
 
\section{Modelling the Condensate Vacuum}
\label{modelling}
Since it is very difficult to treat 
bound states by  the true 
Bethe-Salpeter\cite{BS},
we tend to use instead non-relativistic 
appoximations. In spite of the fact that 
we consider the ``condensate vacuum''
a condensate of these 
bound states, and that they thus 
approximately all should be in the 
same quantum single-particle-state,
we tend to think that the exact 
boson condensate is less important than
that the particles have the mutual 
configuration dimishing the energy density.
We thus suppose that for the  purpose of 
estimating the energy density of such a  
vacuum it is advisable to use rather some 
chrystal model for this vacuum, so that
the way the particles may have arranged 
themselves relative to each other is 
hopefully more realistically taken into 
account. Since the constituents - the 
quarks and anti quarks - are fermions,
we imagine, that the different bound 
states
cannot penetrate into each other more 
deeply than this fermi-statistics allows.
The fermions in one  F neighbor to
an other one  cannot  go deeper in than 
to the level with the main quantum number 
(in the atomic physics language) n=2.
The n=1 level is namely occupied by the 
constituents of the first F.
Now there are for n=2 four orbital states,
one 2s and 4 2p-states, each of which 
in the case of say top 
quarks can contain 2*3 = 6 allowing 
4 *6 =24 top   quarks as 
constituents 
in  the closest neighbor F's to the 
given one. There are  of course also 
similarly place for 24 anti top quarks
in these neighbors. So there is in the 
closest layer of F's around a given one 
place for (24 +24)/12 = 4 neighbors.
Thus the analogy with a material with 
tetra-valent atoms (=having configuration 
number 4) 
such as carbon is called for. Therefore 
diamond is 
a candidate for a model for this 
``condensate vacuum''. Colin Froggatt and
I already used in an appendix in our 
article \cite{Tunguska} such a diamond
model. Via slightly indirect arguments we 
then very crudely reached a 
mass- prediction of 285 GeV for the 
bound state.
In the present 
article the estimate is made a bit more 
directly,  and the result comes a bit 
more like   690 GeV.
    
This analogy-choice means, that we decided
to evaluate the energy density of the 
``condensate'', or perhaps rather a 
chrystal of F particles 
sitting in a diamond shaped lattice 
attrackted most importantly to their 
four nearest neighbors F's. The 
attracktion potential is suggestively 
approximately estimable by considering 
an F neighbor to a given one 
having its 12 top and anti top quarks 
going around the latter in the main 
quantum number orbit n=2. If one could use
pure atomic physics and ignore the Higgs 
mass, it would be wellknown that the 
binding energy in such an n=2 orbit is 
just 1/4 of the binding energy in the 
n=1 orbit, the Rydberg. The binding energy
of one top quark or of one anti top in 
an F is of course, if it consists of 
12 quarks or anti quarks of the same sort
$\frac{12 m_t - m_{F}}{12} 
= \frac{12*173 \; GeV - 750 \; GeV}{12}
=111 \; GeV$ for the case of 
top-quarks with mass $m_t = 173 GeV$
and an ansatz of 750 GeV for the mass 
of F as representing the two slightly
different estimates given below for what 
our earlier work 
suggests\cite{LNvacuumstability}, 710 GeV 
and 850GeV.    
If we ignore the screening effect, which
is claimed to be just compensated for in 
section \ref{accident}, the binding of 
top quark say due to a neighboring F
in the diamond-like lattice is thus 
just 1/4 of this $111$ GeV, i.e. 
it is 28 GeV (really 27.65). 

Our calculation below now consists in 
observing, that such 28 GeV per 
constituent
top or anti top means, that a full F
is bound to a neighbor by 12 * 28 GeV 
= 331 GeV. Now there are four neighbors
but if we use that, we double count by a 
factor 2, and so at the end the binding 
per F-particle in the chrystal 
believed to be an alternative vacuum of 
the type suggested by MPP is 4/2 times 
this 331 GeV. That is to say, that the 
binding between neighbors in the 
diamond-like chrystal per F runs 
up to 4/2 * 331 GeV = 662 GeV. If  this 
is taking into account our very crude 
estimates just equal to a mass 750 GeV 
for the  bound state. Thus in our diamond 
chrystal model for the ``condensate 
vacuum'' we find that the Einstein(/mass) 
energy 
750 GeV of the F in the supposed
vacuum is just cancelled by the binding 
to their neighbors, accurately as our 
multiple point principle (MPP) requires!

This is a coincience, that only happens 
for the mass of the resonance having  the 
right value. But it is of course, if taken
seriously, an evidence for our hypotesis
the Multiple Point Principle. 

Below in section \ref{mass750} we shall 
calculate, what mass 
is it, that is required for this 
cancellation of binding energy against 
the Einstein energy of the 
F-particles thus  allowing the energy 
density of the ``condensate vacuum'' to
be zero.

\section{An Accident Making 
Calculation Easy}
\label{accident}
The most important (attracktional) 
interaction causing the
bound states such a low mass as 
 $\sim$ 750 GeV say 
is in our picture  the exchange 
of Higgs particles, meaning that it is the 
Yukawa potential. Most simply we would 
make an approximation, in which  the top 
and anti top constituents are kept 
together by a Higgs field centered around
the ``center'' of the F-particle (=
the bound state(which {\em we} previously 
called S). Then in order that 
this approximation should be a good one 
for a top or an anti top far away from the 
at (relatively) long distance from
this ``center'', we must make the strength 
of the Higgs Yukawa field around this 
``center'' to be  12 or 11 times that of 
only one top or anti top. However, if at 
shorter
distance we want the true strength of the
Yukawa potential, we should rather only 
use as the strength of this the potential
from this ``center'' being the half  of 
that,
i.e. as the one from rather 11/2 or 12/2
tops or anti tops.
In other words, if we decide to work in 
the crude approximation of the potential 
being simply mathematically a Yukawa 
potential around a ``center'', i.e.
having the form 
\begin{equation}
V(r)= K \exp(-m_{H \; eff} r)/r,
\label{Yukawa}
\end{equation}
  then the constant $K$ should rather 
be proportional to 11/2 than to 11, when
there are 12 constituents in the bound 
state F. Indeed if all the 11 other 
constituents than the one considered were
concentrated at origo we would have 
\begin{equation}
\hbox{With no smaering out: } K = 
\frac{11(g_t/\sqrt{2})^2}{4\pi},
\end{equation}
while, if we, as we must realistically 
take it, have that on the average the 
constituents contributing to the 
attracktion of a given quark or anti quark
are only the half of them being closer to
the center than the given quark itself,
then the Yukawa coupling coefficient 
$K$ should rather be
\begin{equation}
\hbox{In the average distance: } 
K = \frac{11/2 *(g_t/\sqrt{2})^2}{4\pi}.
\end{equation}
This complicates the calculation, 
especially since truly the potential 
inside the bound state is not of the 
simple mathematical form (\ref{Yukawa}),
but rather is a sum over several Yukawa 
potentials, one for each of the 
constituents.

Now, however, we want in the present 
article aiming at a surprisingly simple
``calculation'' to consider a somewhat 
accidental compensation of this reduction 
in the Higgs field strength due to the 
smear out of the center by the constituents
away from the very center by some 
corrections. The corrections which 
Froggatt and I considered in the article 
\cite{4nbs} and which could 
potentially help are:

\begin{itemize}
\item Exchange of gluons 
\item What we call u-channel exchange.
\item Exchange of ``eaten Higgses''.  
\end{itemize}        
  
These are various interactions studied 
by us in the article \cite{4nbs},
which for the case of the two first 
ones at least can in the limit of 
small Higgs mass be included simply 
by replacing the coefficient $K$ of the 
Yukawa potential by a larger value.

In our work \cite{4nbs} we in fact 
correct the simple (t-channel) - which is
one to use, if the interacting quark 
is not exchanging quantum numbers under the Higgs exchange like in a u-channel 
scattering  and the color 
field is screened and other exchanges 
ignored - Yukawa potential 
$V_{t-ch.}(r)= \frac{11/2 *(g_t/\sqrt{2})^2}
{4 \pi *r}$ by replacing it by 
a $V_{total}$,
which is given by
\begin{eqnarray}
V_{total}& =& V_{gluon} + V_{with \;
u−ch},\\
\hbox{where } V_{with \; u-ch.} &=& V_{t-ch.}+ 
V_{u-ch.}(r)= 
-\frac{11/2 * (g_t/\sqrt{2})^2}{4\pi*r}+
-\frac{5/2 * (g_t/\sqrt{2})^2}{4\pi*r}\\
&=&- \frac{16/2 * (g_t/\sqrt{2})^2}{4\pi*r},\\
\hbox{while } V_{gluon}&=& −
\frac{g_s^2 
Tr(\lambda_a /2 ∗ \lambda_a /2)_{\underline{3}}}
{4 \pi Tr(I)_{\underline{3}}r}= 
\frac{g_s^28/2}{4\pi*3r} 
= −e_{t\bar{t}}^2 /(4πr).
\end{eqnarray}

With $g_s^2/(4\pi) = 0.118$ we got 
$e^2_{t\bar{t}}= 1.83$.

This means that the binding energy (4) 
should be corrected to include the gluon 
exchange
force by substituting

\begin{eqnarray}
16g^2_t/4&\rightarrow & e_{tt}^2 + 16g^2_t/4 .\\
\end{eqnarray}
However, even though this compared to 
the t-alone $V_{t-ch}$ corredponding to
$11g_t^2/4$ is already roughly a doubling,
the potential does at first to see not 
change its shape, and thus at first  
the effect of the effective charge being 
distributed, rather than concentrated in 
the center exactly, has not been corrected
for. Thinking a bit more this is , 
however,not quite true because: The 
gluon exchange
is actually quite absent, when we ask for 
the exchange force between two neighboring 
F-particles, because the F's are
from outside seen colorless. So when 
asking for the potential keeping the tops
and anti tops together inside the F-bound 
state, the gluon force is there, but 
when asking for the potential between the 
two F's interacting with each other,
only the Higgs exchange is present. 
This effect gives in fact some 
compensation for the mistake one would do
by ignoring the smearing out of the 
central attracktor (smearing out of 
the analogue of the atomic nucleus in 
atomic physics). 

It is more obvious that 
the third type of correction - ``the eaten 
Higgs exchange'' - is {\em not of the 
same shape} and thus can make a 
difference. The point is that, when an 
``eaten Higgs'',
meaning really exchange of a longitudinal
W or $Z^0$, is exchanged,  the top will 
go into a
different state, such as a left b-quark
state instead. Such a changed state does 
not bind equally strongly as the then 
missing top or anti top would have done.
This means that the modified top or anti 
top being the left bottom quark or 
left anti bottom should rather quickly 
be brought back to become a top or 
anti top, if a large energy increase shall 
be avoided, and we of course think of 
the ground 
state. But some 
amplitude for there being a component 
of the modified top or anti top is 
calculable in second order perturbation 
theory. We may think of this interaction 
expressed just in terms of top and anti 
top effectively taking into account 
the second order perturbation, wherein 
the ``eaten Higgs'' has been exchanged 
{\em twise}. 
Such a double eaten Higgs exchange 
effectively represents,    
 rather than the long range 
Yukawa type interaction, a shorter 
range interaction having rather the  form 
$\propto 1/r^2$ than the Yukawa going 
rather $\propto 1/r$. So the eaten Higgs
correction will change the effective 
shape of the potential. It namely 
contributes mainly only at short 
distances $r$.

At formally very small distances the 
three eaten Higgses couple very similarly 
to the uneaten one, so one should think
it would be like $g_t^2$ were increased 
by a factor 4. But as the distance $r$
gets larger the eaten Higgs attracktion 
will completely disappear. But taking 
it more in conformity with the 
speculations in our work\cite{4nbs} it is 
the combined 
effect of two Higgses (eaten or not)
which goes by a Feynman diagram with 
four $g_t$ factors, that goes up by the 
factor 4 being the number of eaten plus 
uneaten Higses compared to the only one 
uneaten Higgs.  

If we take this correction of the 
coupling square $g_t^2$ to mean that it 
gets increased by 
a fator 2 in the tight region, while there 
is approximately no change  due to the 
eaten 
Higgses for large distance r, then we 
could say: The $g_t^2$ effectively to be 
used for the main quantum number n=1 
should be increased, at most though up 
to being doubled, while the force in the 
 main quantum
number n=2 orbit should be essentially 
unchanged 
by the eaten Higgses. This would 
potentially compensate the effect of 
the only 
11/2 or 12/2 instead of 11 or 12 due to 
the screening, that half the 
attrackting quarks or anti quarks 
are outside the quark, say, to be 
attrackted and thus do not provide any 
attracktion.
However, as just said the factor 2 
increase of the effective $g_t^2$ due
to eaten Higgses were rather an upper 
bound than the most honest estimate.
But now it is then very good for obtaining
an approximate compensation of making 
the error of ignoring the smear out 
of the cloud of the attrackting particles,
that also the gluon exchange gives a 
contribution to the attracktion of the 
tops and anti tops being bound to form the
F, while NOT contributing to the 
attracktion between neighboring F's. 

Now what we are really going to use 
our ``accidental'' cacellation 
of the smear out of the consisuents 
for is to obtain the ratio of the binding 
energy of a constituent in the n=2 orbit 
to be just 1/4 of that in the n=1. Of 
course as long as the effetive Higgs mass 
would be positive, there would allways be
an error in this use because the n=2 
binding energy will be numerically 
suppresed more the effective Higgs mass
than the n=1 orbit. The rudiment of the 
Higgs mass non-zero effect will also 
contribute to counteract to the ignored 
effect of the smearing out of the 
constituents.   

So as a crude, more or less accidental, 
cancellation we shall simplify our 
calculation by taking it, that the 
eaten Higgs correction helped by a couple 
of smaller effects drops out 
against the effect of half the quark 
or antiquark sources being outside 
for the n=1 orbit. This then means 
that we can formally allow ourselves 
to calculate {\em as if} there were 
no sources-outside-effect and no 
eaten Higgs effect, and even as if the 
gluon attracktion had same coupling 
in both n=1 and n=2, being totally 
absorbable in the Higgs exchange. 
So we could go on very simply both 
ignoring gluons - absorbed into Higgs -
and the smearing out of the attrackting 
cloud, being compensated for by other 
effects.

This would mean, that we could calculate 
as if: 
{\em Each quark were bound by a completely
central point Yukawa potential and the 
very same strength could be used then for 
all the orbits n=1 , n=2, and so on.}

That is to say we could claim, that after 
this cancellation - a bit accidentally -
the binding energy of  say a quark 
in the zero Higgs mass limit would be 
just bound by a factor 1/4 in the n=2 
orbit 
compared to  its binding in an  n=1 
orbit.
       
\section{Higgs-field and Higgs 
Effective 
Mass in the Different Vacua}
\label{effHiggsmass}
The zero Higgs mass approximation is, 
however, not so obvious, and if we cannot 
use it, 
  then of course  the binding in 
the 
n=2 orbit might not even be 1/4 of that
in the the n=1 orbit as we claimed 
after our assumed cancellation, see 
section \ref{accident}. 

We already mentioned, that we here would 
make another rather drastic approximation:

We would replace the exponential factor 
in the Yukawa potential by a 
theta-function
like function. That is to say we would
in the different situations, meaning 
respectively 
\begin{itemize}
\item inside the condensate 
of F's, and 
\item in the ordinary/physical 
vacuum, in which we live.
\end{itemize}
use {\em different effective Higgs masses
and thus different theta-like functons
for the exponential factor in the 
Yukawa-potential}

\section{Crude But Easy Treatment of 
the Yukawa Potential Due to 
Higgs Boson}
\label{potentialapproximation}
Since the average Higgs field in the 
condensate vacuum is clearly smaller 
than in the ordinary or physical one, 
the effective Higgs mass will due to 
self-interaction easily be seen to 
be also smaller in the condensate than 
in the ordinary vacuum. 
This is seen from the following:
\begin{itemize}
\item Since say the top quark gets 
its mass from the interaction with the 
vacuum Higgs field, it is obvious 
that a diminishing numerically of the 
Higgs field just around a top quark 
would lower the energy/mass. This 
possible adjustment to lower the energy 
is in fact what brings about the Yukawa 
potential around say a top quark. To 
minimize the energy by adjusting the 
Higgs field around a top it actually 
pays energetically to let the Higgs 
field diverge
infinitesimally near to the top, but 
in the surrounding region the Higgs field 
further and further away go back to its 
usual vacuum expectation value. This 
is how the Yukawa potential comes about.
\item If the Higgs field in some region 
remains very close to some value 
$\phi_0$ say, then small deviations 
in the Higgs field $\phi_H$ from this 
$\phi_0$ 
will behave as if the Higgs mass squared 
were $m_{eff \; Higgs \; mass}^2=
\frac{\partial^2 V_{eff}(\phi_H)}{\partial 
\phi_H^2}|_{\phi_H = \phi_0}$. Now the 
Higgs effective potential actually has 
such a form, that when the value is 
numerically lowered the second derivative 
also becomes lower, end even at some point
becomes zero and then negative. It is 
wellknown that for zero Higgs field 
the effective Higgs mass square in our 
sense 
here is the tachyonic Higgs mass square,
i.e. negative.     
\end{itemize}
That of course 
in turn means that the zero Higgs mass 
approximation gets better
in the condensate than in the ordinary 
vacuum. It is therefore possible
 and we shall assume that - hopefully 
after a numerical estimate, that it 
happens to be so approximately - while 
for the condensate vacuum the zero 
Higgs mass can be used {\em including}
the n=2 orbit, i.e. for orbit n=1 
{\em and} 
n=2, in the ordinary vacuum only 
the n=1 orbit allows the zero Higgs 
approximation, for the higher ones we 
let the Yukawa potential completely 
be approximated 
by zero(this corresponds to infite 
Higgs mass relative to the inverse radius 
for these higher than or equal to n=3
orbits).     
 
The here suggested treatment of the 
Yukawa potential is not exactly to take it
as a step function, but it is 
approximately
so. 

To summarise the rule suggested - and to 
be confirmed by some estimations - : 
\begin{itemize}
\item{Condensate vacuum} In the 
condensate vacuum we take the Higgs mass
zero for orbits n=1, and 2, and huge, say 
infinite, for n=3,4,... 
\item{Present vacuum} while in the 
vacuum, we live in, we instead take
Higgs mass zero (only) for n=1, while 
infinite or huge for n=2,3,4,... 
\end{itemize}

\section{Calculation of the Mass of 
the F Bound State}
\label{mass750}
The basic requirement used in the present
``calculation'' of the mass of 
bound state S=F of 6 tops + 6 anti 
tops
is, that in a presumably too naive 
nonrelativistic thinking the energy 
density of the condesate vacuum shall
be zero. Of course this zero 
should be understood as being compared 
to the background being
identified with the vacuum, in which we 
live. Taking the zero to be relative 
to energy density in the usual vacuum, 
means that we only have to include those 
energy carying ingredients, which are not 
also present in the usual vacuum, in
which we live. Since the ``condensate 
vacuum'' is characterized by its extra 
F-particles, it is the energy density
resulting from these particles and their 
mutual interactions, that should be added
up and required by MPP to be zero.   

That is to say we shall assume, that the 
energy in the condensate vacuum per
F-particle, i.e. really 
$\rho_{F}^{-1}\rho_{energy}$, is zero. 
Here we used the notation that 
$\rho_{F}$ 
is the density inside the condensate of 
the 
particles F, while the energy density 
of the condensate, counted relative to 
the vacuum, we live in, and which 
have no F's in first thinking (of 
course
there are some vacuum fluctuations, but 
that is not counted into neither 
$\rho_{F}$
nor $\rho_{energy}$, if the same is 
present in 
the vacuum, we live in.)
 
This MPP-requirement 
is written
\begin{eqnarray}
0 &=& m_{F} - ``\hbox{binding 
per F}''\\
  &=& m_{F}-
\frac{\#\hbox{neighbors}}{2}
* ``\hbox{binding to neighbor F}''\\
&\approx& m_S - \frac{4}{2} *
``\hbox{binding 
of F in n=2 arround another 
F}''\\
&\approx& m_{F} - \frac{4}{2}*
``\hbox{binding of F}''*
\frac{1/2^2}{1/1^2}\\
&=& m_{F}-\frac{1}{2}* 
``\hbox{binding of F}''\\
&=& m_{F} -\frac{1}{2}*(12m_t-m_{F})\\
&=& \frac{3}{2}m_{F} - 6 m_t 
\label{zerocond}
\end{eqnarray}
We shall indeed follow an appendix of 
our earlier work\cite{Tunguska} and 
assume,
that the structure of the condensate can 
be approximated as being a diamond lattice
structure, so that there are just
$\#\hbox{neighbors} = 4$, i.e. other 
F-particles 
surrounding each one of them in the 
lattice. When we count all the binding 
energy per F present in the 
condensate
$``\hbox{binding per F}''$ as being 
the 
number of neighbors $\#\hbox{neighbors}
$ times   the binding 
of one F to its neighbor 
$``\hbox{binding to neighbor F}''$,
we {\em double count}, because we 
count the same binding from both the 
one F and from the other one it 
binds. 
Therefore we must have the denominator $2$
seen in the formula (\ref{zerocond}).

We made then the approximation, 
that we can effectively consider it, 
that the neighboring top quarks and 
anti topquarks contained in an F 
neighboring to another one are {\em in 
effect 
in the n=2 orbit} of the latter. Thus we 
can take the binding energy of a 
neighboring F to a given one   
$``\hbox{binding to neighbor F}''$
to be  as, if the top and anti tops 
were in an n=2 orbit or some 
superposition thereof. Thus the binding
of the neighbors occur with binding energy
$``\hbox{binding 
of F in n=2 around another F}''$.

As long as we can take the effective 
Higgs mass for the two lowest orbits 
n = 1 and 2 to be zero, we can count, that
the binding energy, for top say, in the 
orbit n=2 is just one quarter of that 
in the n=1 orbit, provided we can use the 
same potential of the form $\propto 1/r$.
But now that were, what our above 
discussion 
``accidental cancellation'' in 
section \ref{accident} should ensure, and 
so even for an F-particle,
 which consists 
of tops and anti tops the ratio of the 
binding energies should be $1/2^2 = 1/4$.

From the last step in (\ref{zerocond})
we easily derive of course
\begin{equation}
m_{F} = \frac{2}{3}*6m_t = 4m_t 
= 173 GeV 
*4 = 692 GeV \hbox{agreeing well with } 710 GeV \hbox{ or } 850 GeV!
\end{equation}

\section{Calculation of the 
Mass of the into 
Gauge 
Bosons Decaying Observed (?) 
1.8 TeV-Resonance}
\label{mass1c8}
In general our model means that
the Higgs coupling of our bound 
state, F or S, is very strong, 
and several such bound states
interact - actually attrackt -
each other. Thus we expect that 
there should exist further 
resonaces formed from two or more
bound states. After all we even 
have used the picture that there 
is new phase a new vacuum, which
might be considered a huge 
bound state of an infinite number
of our bound state of 6 top and 6 anti top, F. The most significant
of such further bound states of 
our bound states would presumably
be a resonance of just two of 
them.

It happens that    
the resonance doubtfully 
seen\cite{r18} 
in the 
decay into 
Z's or W's with mass 1.8 TeV 
could 
be identified with a resonance 
consisting
essentially of two F-particles, 
which 
in turn are the bound states of 
the 
6 tops and 6 anti tops each. 

Already the fact that the 
mass 1.8 TeV 
is very crudely just that of two 
F-particles(with mass as our 
bound state
calculation based on MPP 
suggests), 2* 750 GeV = 1500 GeV 
= 1.5 TeV, suggests such a 
thinking. But 
now we shall estimate the mass 
of the 
1.8 TeV resonance by the 
following  
proceedure:

{\em If} we could calculate as above that 
the attracktion in the n=1 and n=2 
orbits were as if the Higgs mass 
were zero, we could have one F 
bound to 
another one by having all the tops and 
anti tops of the one going into the n=2 
orbit of the other one. Then the binding 
 would  according to our rule 
used above be just 1/4 of the binding 
of one F-particle $12m_t - m_{F}$. Now,
however, because of the effective Higgs 
mass being smaller in the situation, 
wherein we have only two F-particles 
rather than a full condensate, as we 
thought about above, we should take it
that the attracktion in the n=2 disappears,
when going from the condensate to the 
system - the 1.8 particle we hope - of 
only two F's.
This change we then treat as a 
perturbative 
correction to the first mass for the 
two F-bound state of 
$2m_{F} - \frac{1}{4}*(12m_t-m_{F})$ due to 
the change 
of the potential energy between the 
two F-particles disappearing. 

Now in potentials of the $1/r$ form,
which we use as our approximation 
here, the virial theorem allows us 
to use, that the pototential energy
is negative and just twise the 
binding energy. The energy or mass 
of the two-F-resonance will thus 
rather be {\em increased} compared to 
the mass of two F-particles than 
decreased
by the amount of $\frac{1}{4}
*(12m_t-m_{F})$. 

That is to say the mass of the resonance 
- which we want to identify with the 
1.8 TeV finding - is given as
\begin{equation}
m_{2Fresonance} = 2m_{F}  + 
\frac{1}{4}
*(12m_t-m_{F}) = 1.75m_{F}+3m_t = 1313 GeV
+519 GeV = 1832 GeV = 1.83 TeV,
\end{equation} 
which agrees - accidentally? - wonderfully
with the number 1.8 TeV from the 
experiment! 
\subsection{Excuse for Narrow 
Width}
\label{Width}
When you think about our huge Higgs mass 
approximation for the case of binding 
of just two F-particles, you see that 
formally there is no interaction at all 
between the two! This would not allow a 
narrow resonance. Therefore we need 
some helping story to excuse that the 
width after all gets so small, that we 
experimentally shall conceive of the 
resonance with mass 1.8 TeV as a 
resonance at all. 

The excuse suggested is, that during the 
interaction assumed to be mainly given by 
Higgs exchange in our calculation the two
6tops + 6 anti tops boundstates F 
are  {\em 
partly annihilated} so as to be in 
reality 
not really two true F-particles, but 
rather some similar structures with 
a bit smaller numbers of constituents.
We could still hope that although such 
a partial annihilations, that could go 
back and forth, while the two F's 
move 
arround, could help to decrease the 
decay rate and making the width of the 
1.8 TeV-resonance smaller, it would not 
severely modify our crude estimate above.

The narower we can speculate the 1.8 TeV 
resonance to be w.r.t. decaying into
two F's the bigger we can hope for 
the 
partial width into other channels than 
the channel into two F's. Since it is 
so far only seen in vector boson channels, 
it is needed to be speculated, that the 
two F's channel does not take away 
all 
the 1.8 TeV particles.      

\section{Conclusion}
\label{conclusion}
\subsection{The New ``Calculation''}
We have presented an overly 
simple 
``calculation'' leading to 
there being 
due to the strong top-Yukawa 
coupling 
 and under the assumtion of a 
finetuning 
ensuring a with the normal vacuum 
degenerate one with a 
condensate of bound 
states S=F of six top + six 
anti top, two 
``resonances'' with masses 
respectively 
\begin{eqnarray}
m_{F} & = & \frac{12 m_t}{3} = 4 m_t= 
692 GeV\\
m_{1.8} & = & 2 m_{F} 
+\hbox{ ``kinetic energy of n=2 orbit''} = 2m_{F} + 
\frac{12m_t}{6}
= 2m_{F} + 2m_t =1846 GeV  
\end{eqnarray}
These results were obtained in 
the 
philosophy, that the coupling constants 
- especially say the top-Yukawa-coupling 
$g_t$
- are by the new principle ``multiple 
point principle'' adjusted/finetuned 
to make 
the energy density 
of a condensate of 
F-particles - which {\em we} 
earlier caled 
S-particles -
 (the bound states of 6 top 
+ 6 anti top quarks) have just 
the same energy density as the 
usual 
vacuum (in which we live). That 
were 
to say,
that the interaction between the 
F-particles in the condensate 
 should 
be 
just so strong an attracktion between 
them, that the total energy (density) of 
the 
condensate just becomes zero (relative 
to the normal vacuum). That is to say the 
binding between the neighboring 
F-states 
just equals the mass-energies of these 
F-states. If one therefore considers 
our good agreement of the masses as an 
evidence for the truth of the assumptions
having been used, then there is a 
significant evidence for our long 
speculated ``Multiple Point Principle''!

It must, however, be admitted that the 
present very simplified ``calculation''
were based on a very crude treatment 
of the Yukawa potential representing the 
Higgs exchange between the 
F-particles,
which is the (main) interaction between
these F-particles in our picture. 
In fact we approximated 
Higgs exchange by letting the Higgs be 
effectively massless, when the top-antitop
quarks are in relative orbits with 
atomic main quantum number n=1, while 
we let the Higgs potential be  
totally cut away for n=3, 4, ... . For the 
main quantum number n=2 we made the more
complicated assumption of letting it
be either as for a zero mass Higgs or 
cut down to zero according to the 
surroundings, which influence the 
average Higgs field. Indeed we took 
for n=2 the Higgs exchange force to be 
like for massless Higgs inside 
the 
condensate vacuum, while we 
put the 
exchange potential to be zero, 
when 
applied in the two-F-system 
identified
with 
resonance hoped for with the mass
 1.8 TeV. Although the resonance 
previously experimentally 
suggested as  
excess seen decaying into 
weak gauge bosons does not seem 
much supported, it would for 
our system be very fine with such
an 1.8 TeV resonance. In general 
some related resonances to our 
main bound state with mass in 
the 750 GeV range are in our 
scheme not unexpected, since we 
after all have a scheme with 
- a new type(i.e. not just
QCD.) of - strong 
interactions. The mass region at
about 1.8 TeV is the first 
suggested such further resonance
to be expected.  
 
Thus more severe calculations are to 
be performed to truly settle, if our 
calculations are right. It should in this 
connection be stressed, that since our 
model
is in principle {\em only Standard Model}
extended with our Multiple Point Principle 
used to restrict the coupling constants,
one should in principle be able to  
calculate whatever one wants. With a 
relatively strong coupling $g_t$ being 
the very basis for the whole story there 
is though of course the complication 
of not having in principle the basis for 
perturbation theory.
\subsection{Main Coincidence!
Earlier Bound State Mass 
Fit from MPP}

\label{Larisaetal}
It should be stressed that the 
main point and result of our 
estimate that our mass estimate 
{\em coincides with an earlier 
result obtained also using MPP
but using a differeent vacuum
the ``high Higgs field vacuum''}
- we could call it-:  

A recent work by Larisa Laperashvili,
 Chitta Das and 
myself\cite{LNvacuumstability}, in which 
we have the bound state, 
F-resonance, give a little 
correction 
to the mass of the Higgs, that should be 
measured relative to the one associated 
with the running self coupling at the weak 
scale, improves the agreement with 
exoeriment of requirement of 
Multiple Point Principle for yet a vacuum.
Indeed we have in our picture a third 
vacuum (in addition to the usual one and 
the condensate vacuum with  its F's
in it), namely one   with a very high 
Higgs field 
expection value. According to 
Standard Model caculations witout 
our bound state the high Higgs field 
vacuum has a slightly {\em negative}
energy density (compared to the two 
other vacua). However, we find 
a little correction depending among on 
other quantities on the mass of the 
bound state. We found that this 
bound state mass put to 750 GeV 
would fit wonderfully and 
consider,
that this fact  strongly supports the 
truth of  MPP. In fact it turns out 
that the mass 750 GeV for the bound state
F is perfect for the correction of 
ours just to bring the energy density 
of the vacuum with the high Higgs field 
expectation value from its otherwise 
slightly negative value to zero. This 
means that our Multiple Point Pinciple 
using a quite third vacuum, namely one 
with a high Higgs field expectation value
of the order of $10^{18}$ GeV, leads to 
a need for a particle - our bound state
indeed - with a mass about  750 GeV
also. This means that now , when the F(750)
fluctuation digamma once so fashionable 
turned out  being only a statistical 
fluctuation, then we would nevertheless
from Multiple Point Principle get {\em 
two different and essentailly independent 
reasons for our  bound state to have the 
mass near the value 750 GeV}. That is to
say we would then claim, well our bound 
state should be with a mass 
close to 750 GeV. We have already two 
calculations of this mass in different 
ways, although both originating from the 
same principle MPP, but involving quite 
different data to fit.

In fact Larisa Laperashvili et al.
\cite{LNvacuumstability} uses a corection
due to the diagram
\begin{center} 
\includegraphics{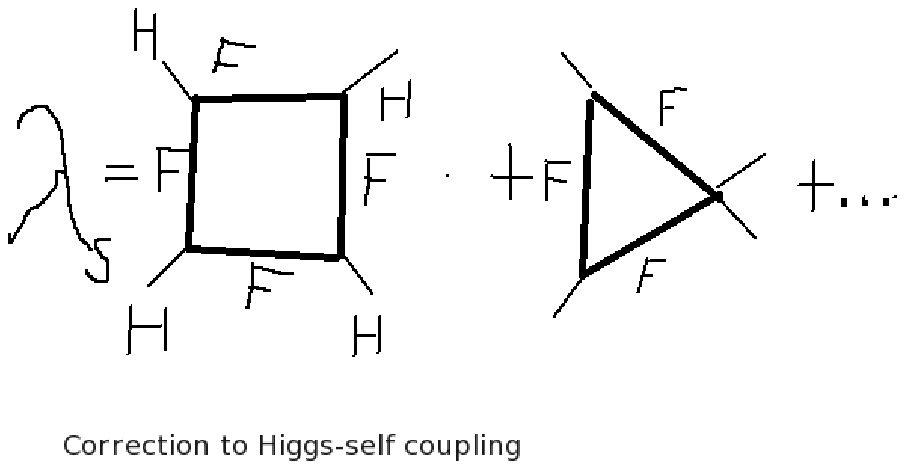}
\end{center}
\vspace{-20 mm}
in order to correct the running self 
coupling of the Higgs $\lambda_{run}(10^{18}
GeV)$ corresponding to the observed Higgs 
mass 125 GeV to go from the value obtained 
by DeGrassi et al. \cite{Deg} of
\begin{equation}
\lambda(\phi_{\hbox{``high field''}}) = -0.01 \pm 0.002.
\end{equation}
at the high field scale 
$\phi_{\hbox{``high field''}}$ 
to the value very accurately zero 
requied by Multiple Point Principle(=MPP).
Since bound state F 
is an extended object we must 
include a formfactor, when using it.
 
Defining a quantity $b$ denoting the 
radius of the bound state measured with 
top quark Compton wave length $1/m_t$ as 
unit by:
\begin{eqnarray}
<\vec{r}^2>& =& 3r_0^2,\\
r_0 &=& \frac{b}{m_t},
\end{eqnarray}
 we obtain a theoretical estimate 
\begin{equation} 
b=\sqrt{\frac{<\vec{r}^2>}{3}} m_t \approx 2.34, \label{9w} 
\end{equation}
crudely confirmed by a slightly different 
estimate.

The dominant diagram/correction - the 
first and quadratic of the diagrams on 
the figure just 
above - is 
$$ \lambda_S \approx 
\frac{1}{\pi^2}\left(\frac{6g_t}{b} \frac{m_t}{m_S}\right)^4$$
where we have the estimated or measured 
values 
$$ g_t =0.935 ;\;  m_t = 173 GeV;\;
b\approx 2.34 \hbox{or} 2.43$$

Using the after all rather small deviation 
from perfect MPP 
$$\lambda_{\hbox{high field}} = -0.01 \pm 0.002$$  and requiring it to be cancelled by 
the correction from the bound state we 
get the requirement
\begin{equation}
\lambda_S = \frac{1}{\pi^2}\left ( 
\frac{6g_t}{b}*\frac{m_t}{m_{F}}\right )^4 *
( \sim 2) \approx 0.01 \pm 0.002,
\end{equation} 
where $g_t = .935$, $m_t=173 GeV$, 
$b\approx 2.43$ and the factor 
``$(\sim 2)$'' were taken in to 
approximate some neglected diagrams,
as the next on the figure. 
If a nearer study should show that 
the next diagrams add up to roughly as 
much as the first one should include the 
factor $\sim 2$ to take into account
the neglected Feynman diagrams correcting 
the Higgs self coupling.

The solution w.r.t. the mass of the 
bound state $m_{F(750)}$ gives
\begin{eqnarray}
m_{F} &\approx& \frac{6g_tm_t}{b}
\left (
\frac{\sim 2}{\pi^2* 0.01\pm 0.002}\right)
^{1/4} \nonumber\\
\approx 2.31*173 GeV
*2.1& =&4.9 *173 GeV =850 GeV \pm 20 \%
\nonumber\\
\hbox{or without the $\sim 2$:}\;  
m_{F}= 2.31 *173 GeV * 1.8&=&
4.1 *173 GeV =710 GeV \pm 20 \%
\nonumber  
\end{eqnarray} 

The ``without the $\sim 2$ '' means 
what one shall do if the first diagram 
indeed dominates strongly.
\begin{center}
\end{center}

In this way we got even two calculations 
for the bound state mass - using 
in addition crude estimation -
\begin{eqnarray}
m_{F}(\hbox{from ``high field vacuum''})
&\approx & 850 GeV\pm 30 \% 
\hbox{with $\sim 2$}\\
m_{F}(\hbox{from ``high field vacuum''})
&\approx & 710 GeV\pm 30 \% 
\hbox{without $\sim 2$}\\
m_{F}(\hbox{``condensate vac.''})&\approx &
692 GeV\pm 40 \%
\end{eqnarray}
  
The agreement of the value ``692 GeV''
with the estimate(s) from the 
completly different vacuum with 
the high Higgs field ``850GeV'' 
or ``710 GeV'' is 
encourraging and a support of our 
``Multiple Point Principle''!

\subsection{No Genuine New 
Physics}
If our numbers are taken as so convincing
that our picute should be taken seriously
then we would have the consequences:
\begin{itemize}
\item We must take our ``Multiple Point 
Principle'' as a true new physical law, 
even the mechanism behind it may not be 
clear. 
\item We must accept that the Standard 
Model except for ``smaller'' deviations,
that are too small to significantly 
modify the running of the Higgs self
coupling, is valid all the way to about 
$10^{18} $ GeV. Some see-saw neutrinos 
may be acceptable, as long as they do not
couple too strongly to the Higgs to 
influence the running of its sef coupling.
Otherwise it would be very accidentally 
that a pure Standard model calculation 
would give so consisten results. 

So there would not be much place for new
physics except for the various resonances 
formed from the boud states, because we 
now  have a new regime of strong 
interactions
that can only be treated by 
non-perturbative methods. (The suggestion 
for the 
1.8 TeV resonance is an example for how 
there can be more particles in such 
a new strong interaction regime)  
\item There should be seen sooner or laler
 bound state of the 6 top + 6 anti top 
with a mass not far from 800 GeV.
\end{itemize}

But even if it thus looks a bit sad w.r.t.
much new physics, one should not forget 
that having our Multiple Point Principle 
established would be a strong element 
of new physics, perhaps then of an 
a bit unexpected type.
         
\section*{Acknowledgement}
I want to thank the Niels Bohr Institute 
for allowing me to stay as emeritus. 
Also I thank my near collaborators on 
the subject of the present article 
Larisa Laperashvili, Colin D. Froggatt
and Chitta Das for the collaboration, 
without which the present article 
could never have come to being.
Further I thank for the comments I have 
got at Mittag Lefler Insitute and the 
Bled workshop ``Beyond the Standard 
Models'' where I have had some 
presentation of essentially the present 
work.   

Astri Kleppe even presented our 
``Mutiple Point Principle'' for the 
workshop.


\begin{thebibliography}{99}
\bibitem{FN750} 
  C.~D.~Froggatt and H.~B.~Nielsen,
  ``Production and Decay of 750 Gev state of 6 top and 6 anti top quarks,''
  arXiv:1605.03909 [hep-ph].


\bibitem{LNvacuumstability}
  L.~V.~Laperashvili, H.~B.~Nielsen and C.~R.~Das,
  ``New results at LHC confirming the 
vacuum stability and Multiple Point 
Principle,''
  Int.\ J.\ Mod.\ Phys.\ A {\bf 31} (2016) no.08,  1650029
  doi:10.1142/S0217751X16500299
  [arXiv:1601.03231 [hep-ph]].
\bibitem{Kawana1}
Y. Hamada, H. Kawai and K. Kawana, “Evidence of the Big Fix,” Int. J. Mod. Phys. A 29, no. 17, 1450099 (2014)
[arXiv:1405.1310 [hep-ph]].
\bibitem{Kawana2}
 Y. Hamada, H. Kawai and K. Kawana, “Weak Scale From the Maximum Entropy Principle,” arXiv:1409.6508
[hep-ph]
\bibitem{5mp} D.L.~Bennett and H.B.~Nielsen, Int. J. Mod. Phys. A{\bf 9}, 5155
(1994).
\bibitem{6mp} D.L.~Bennett and H.B.~Nielsen, Int. J. Mod. Phys. A{\bf 14}, 3313
(1999).
\bibitem{7mp} D.L.~Bennett, C.D.~Froggatt and H.B.~Nielsen, in Proc. 27th
Int. Conf. on High En- ergy Physics, Glasgow, Scotland, 1994, eds.
P. Bussey and I. Knowles (IOP Publishing Ltd., 1995), p. 557.
\bibitem{8mp} D.L.~Bennett, C.D.~Froggatt and H.B.~Nielsen, in Perspectives
in Particle Physics 94, eds. D. Klabu.car, I. Picek and D. Tadi.c
(World Scientific, Singapore, 1995), p. 255.
\bibitem{9mp} C.D.~Froggatt and H.B.~Nielsen, Phys. Lett. B{\bf 368}, 96 (1996).
\bibitem{10mp} L.V.~Laperashvili, Yad. Fiz. {\bf 57}, 501 (1994) [Phys. Atom. Nucl.
{\bf 57}, 471 (1994)].
\bibitem{11mp} C.D.~Froggatt, L.V.~Laperashvili, R.B.~Nevzorov and
H.B.~ Nielsen, Phys. Atom. Nucl. {\bf 67}, 582 (2004) [Yad. Fiz.
{\bf 67}, 601 (2004)], arXiv:hep-ph/0310127.
\bibitem{12mp} C.D.~Froggatt, L.V.~Laperashvili and H.B.~
Nielsen, Phys. Atom. Nucl. {\bf 69}, 67 (2006), hep-ph/0407102.
\bibitem{13mp} D.L.~Bennett, L.V.~Laperashvili and H.B.~Nielsen, {\it Relation
between fine structure constants at the Planck scale from multiple
point principle}, in Proc. 9th Workshop: What Comes Beyond the
Standard Models, Bled, Slovenia, eds. M. Breskvar et al. (DMFA,
Zaloznistvo, Ljubljana, 2006), p. 10, arXiv:hep-ph/0612250.
\bibitem{14mp} D.L.~Bennett, L.V.~Laperashvili and H.B.~Nielsen,
{\it Finestructure constants at the Planck scale from multiple
point principle}, in Proc. 10th Workshop on What Comes Beyond the
Standard Model, Bled, Slovenia, 1727 Jul 2007, arXiv:0711.4681.
\bibitem{15mp} C.D.~Froggatt, R.B.~Nevzorov and H.B.~Nielsen, {\it Smallness of the
cosmological constant and the multiple point principle}, J. Phys.
Conf. Ser. {\bf 110}, 072012 (2008), arXiv:0708.2907.
\bibitem{16mp}
C.D.~Froggatt, R.B.~Nevzorov, H.B.~Nielsen, A.W.~Thomas, Phys.
Lett. B{\bf 737}, 167 (2014), arXiv:1403.1001.
\bibitem{17mp}
C.D.~Froggatt, R.B.~Nevzorov, H.B.~Nielsen, A.W.~Thomas, {\it On
the smallness of the cosmological constant in SUGRA models with
Planck scale SUSY breaking and degenerate vacua}, 2015 European
Physical Society Conference on High Energy Physics (EPS-HEP 2015),
22-29 Jul, 2015, Vienna, Austria; arXiv:1510.05379.
\bibitem{deriving}
  H.~B.~Nielsen and M.~Ninomiya,
  ``Degenerate vacua from unification of second law of thermodynamics with other 
laws,''
  Bled Workshops Phys.\  {\bf 12} (2011) no.2,  199
  [hep-th/0701018].
\bibitem{Stillits} Stillits, Cand. Scient.
thesis at the Niels Bohr Institute

\bibitem{Tunguska}
  C.~D.~Froggatt and H.~B.~Nielsen,
  ``Tunguska Dark Matter Ball,''
  Int.\ J.\ Mod.\ Phys.\ A {\bf 30} (2015) no.13,  1550066
  doi:10.1142/S0217751X15500669
  [arXiv:1403.7177 [hep-ph]].



\bibitem{FN} C.D.~Froggatt and H.B.~Nielsen, {\it Origin of Symmetries},
World Scientific, Singapore, 1991.
\bibitem{DL} C.R.~Das and L.V.~Laperashvili, Int. J. Mod. Phys. A{\bf 20}, 5911 (2005).
\bibitem{17vac} N.~Cabibbo, L.~Maiani, G.~Parisi and R.~Petronzio, Nucl. Phys.
B{\bf 158}, 295 (1979).
\bibitem{18vac} P.Q.~Hung, Phys. Rev. Lett. {\bf 42}, 873 (1979).
\bibitem{19vac} R.A.~Flores, M.~Sher, Phys. Rev. D{\bf 27}, 1679 (1983).
\bibitem{20vac} M.~Lindner, Z. Phys. {\bf 31}, 295 (1986).
\bibitem{21vac} D.L.~Bennett, H.B.~Nielsen and I.~Picek, Phys. Lett. B{\bf
208}, 275 (1988).
\bibitem{22vac} M.~Sher, Phys. Rept. {\bf 179}, 273 (1989).
\bibitem{23vac} M.~Lindner, M.~Sher, H.W.~Zaglauer, Phys. Lett. B{\bf 228},
139 (1989).
\bibitem{24vac} P.B.~Arnold, Phys. Rev. D{\bf 40}, 613 (1989).
\bibitem{25vac}  G.~Anderson, Phys. Lett. B{\bf 243}, 265 (1990).
\bibitem{26vac} P.~Arnold and S.~Vokos, Phys. Rev. D{\bf 44}, 3620 (1991).
\bibitem{27vac} C.~Ford, D.R.T.~Jones, P.W.~Stephenson, M.B.~Einhorn, Nucl. Phys.
B{\bf 395}, 17 (1993).
\bibitem{28vac} M.~Sher, Phys. Lett. B{\bf 317}, 159 (1993).
\bibitem{29vac} G.~Altarelli and G.~Isidori, Phys. Lett. B{\bf 337}, 141 (1994).
\bibitem{30vac} J.A.~Casas, J.R.~Espinosa, M.~Quiros, Phys. Lett. B{\bf 342}, 171
(1995).
\bibitem{31vac} J.R.~Espinosa, M.~Quiros, Phys. Lett. B{\bf 353}, 257 (1995).
\bibitem{32vac} J.A.~Casas, J.R.~Espinosa, M.~Quiros, Phys. Lett. B{\bf 382}, 374
(1996).
\bibitem{33vac} B.~Schrempp and M.~Wimmer, Prog. Part. Nucl. Phys. {\bf 37}, 1 (1996).
\bibitem{34vac}
 C.D.~Froggatt, H.B.~Nielsen, and Y.~Takanishi,
{\it Standard model Higgs boson mass from borderline metastability
of the vacuum}, Phys. Rev. D{\bf 64}, 113014 (2001).
\bibitem{35vac} V.~Brancina and E.~Messina, Phys. Rev. Lett. {\bf 111}, 241801
(2013), arXiv:1307.5193.
\bibitem{36vac} V.~Brancina, E.~Messina and M.~Sher, Phys. Rev. D{\bf 91}, 013003 (2015),
arXiv:1408.5302.
\bibitem{37vac} V.~Brancina, E.~Messina and A.~Platania, JHEP {\bf 1409}, 182 (2014),
arXiv:1407.4112.
\bibitem{Deg} G.~Degrassi, S.~Di~Vita, J.~Elias-Miro, J.R.~Espinosa,
G.F.~Giudice, G.~Isidori and A.~Strumia, JHEP {\bf 1208}, 098
(2012); arXiv:1205.6497.
\bibitem{But}
D.~Buttazzo, G.~Degrassi, P.P.~Giardino, G.F.~Giudice, F.~Salab,
A.~Salvio,\\ A.~Strumia,  JHEP {\bf 1312},  089 (2013);
arXiv:1307.3536.
\bibitem{1he} G.~Isidori, G.~Ridolfi, A.~Strumia, Nucl. Phys. B{\bf 609}, 387
(2001).
\bibitem{2he} J.R.~Espinosa, G.F.~Giudice and A.~Riotto, JCAP 0805 (2008) 002.
\bibitem{3he} J.~Ellis, J.R.~Espinosa, G.F.~Giudice, A.~Hoecker and A.~Riotto,
Phys. Lett. B{\bf 679}, 369 (2009).
\bibitem{4he} J.~Elias-Miro, J.R.~Espinosa, G.F.~Giudice, G.~Isidori, A.~Riotto,
A.~Strumia, Phys. Lett. B{\bf 709}, 222 (2012).
\bibitem{1nbs}
C.D.~Froggatt and H.B.~Nielsen, {\it Trying to understand the
Standard Model parameters}. Invited talk by H.B.~Nielsen at the
"XXXI ITEP Winter School of Physics", Moscow, Russia, 18-26
February, 2003; Surveys High Energy Phys. {\bf 18},  55-75 (2003);
hep-ph/0308144.
\bibitem{2nbs}
C.D.~Froggatt, H.B.~Nielsen and L.V.~Laperashvili, {\it
Hierarchy-problem and a bound state of 6 t and 6 anti-t,} in:
Proceedings of Coral Gables Conference on Launching of Belle
Epoque in High-Energy Physics and Cosmology {\it (CG 2003)}, Ft.
Lauderdale, Florida, 17-21 December, 2003.
\bibitem{3nbs}
C.D.~Froggatt, H.B.~Nielsen and L.V.~Laperashvili, Int. J. Mod.
Phys. A {\bf 20}, 1268 (2005); hep-ph/0406110.
\bibitem{4nbs}
C.D.~Froggatt and H.B.~Nielsen, Phys. Rev. D {\bf 80}, 034033
(2009); arXiv:0811.2089.



\bibitem{5nbs}
C.D. Froggatt, H.B. Nielsen, {\it Hierarchy Problem and a New
Bound State}, in Proc. to the Euroconference on Symmetries Beyond
the Standard Model, p.73, Slovenia, Portoroz, 2003 (DMFA,
Zaloznistvo, 2003); ArXiv: hep-ph/0312218.
\bibitem{6nbs}
C.D.~Froggatt, {\it The Hierarchy problem and an exotic bound
state,} in: Proceedings of 10th International Symposium on
Particles, Strings and Cosmology, \\ {\it (PASCOS 04)}, Boston,
Massachusetts, 16-22 Aug, 2004. Published in: ``Boston 2004,
Particles, strings and cosmology'', pp.325-334; hep-ph/0412337.
\bibitem{7nbs} C.D.~Froggatt, L.V.~Laperashvili and H.B.~Nielsen,
{\it A New bound state 6t + 6 anti-t and the fundamental-weak
scale hierarchy in the Standard Model,} in: Proceedings of 13th
International Seminar on High-Energy Physics: {\it QUARKS-2004},
Pushkinskie Gory, Russia, 24-30 May, 2004; hep-ph/0410243.
\bibitem{8nbs}
C.D.~Froggatt, L.V.~Laperashvili and H.B.~Nielsen, Phys. Atom.
Nucl. {\bf 69}, 67 (2006) [Yad. Fiz. {\bf 69}, 3 (2006)];
hep-ph/0407102.
\bibitem{9nbs}C.D.~Froggatt, L.V.~Laperashvili, R.B.~Nevzorov and
H.B.~Nielsen, {\it The Production of $6t+6\bar t$ bound state at
colliders}. A talk given  by H.B.~Nielsen at CERN, 2008, preprint
CERN-PH-TH/2008-051.
\bibitem{10nbs}
C.D.~Froggatt, L.V.~Laperashvili, R.B.~Nevzorov, H.B.~Nielsen and
C.R.~Das, {\it New Bound States of Top-anti-Top Quarks and T-balls
Production at Colliders (Tevatron, LHC, etc.)}, Report
CHEP-PKU-1-04-2008, CERN (2008); arXiv:0804.4506.
\bibitem{11nbs}
C.D.~Froggatt and H.B.~Nielsen, in: Proceedings to the 34th
International Conference on High Energy Physics (ICHEP 2008), 30
Jul - 5 Aug 2008, Philadelphia, Pennsylvania; arXiv:0810.0475.
\bibitem{12nbs}
C.R.~Das, C.D.~Froggatt, L.V.~Laperashvili and H.B.~Nielsen, Int.
J. Mod. Phys. A {\bf 26}, 2503 (2011); arXiv:0812.0828.
\bibitem{13nbs}
C.D.~Froggatt, C.R.~Das, L.V.~Laperashvili and H.B.~Nielsen, {\it
New indications of the existence of the 6 top-anti-top quark bound
states in the LHC experiments}, a talk by L.V.~Laperashvili at the
Conference "Quarkonium-2012", Moscow, Russia, MEPhI, November, 12
- 16, 2012, arXiv:1212.2168 [hep-ph]; Yad. Fiz., {\bf 76}, 172
(2013).
\bibitem{14nbs}
C.D.~Froggatt, C.R.~Das, L.V.~Laperashvili, H.B.~Nielsen, Int. J.
Mod. Phys. A{\bf 30}, 1550132 (2015); arXiv:1501.00139.
\bibitem{ATLAS1}
LHC seminar "ATLAS and CMS physics results from Run 2", talks by
Jim Olsen and Marumi Kado, CERN, 15 Dec. 2015.
http//indico.cern.ch/event/442432/.
\bibitem{ATLAS2}
ATLAS Collaboration, ATLAS-CONF-2015-081, "Search for resonances
decaying to photon pairs in 3.2 $fb^{-1}$ of pp collisions at
$\sqrt s$ = 13TeV with the ATLAS detector".
\bibitem{CMS}
CMS Collaboration, CMS PAS EXO- 15-004, "Search for new physics in
high mass diphoton events in proton-proton collisions at 13TeV";
http//indico.cern.ch/event/442432/.
\bibitem{ATLAS}
ATLAS Collaboration (G. Aad et al.), Phys. Rev. Lett. {\bf 114},
081802 (2015).
\bibitem{PDG}
Particle Data Group Collaboration (K.A. Olive (Minnesota U.) et
al.), 2014, 1676 pp. Chin. Phys. C{\bf 38}, 090001 (2014).
\bibitem{Mtexp} ATLAS, CMS, D0 Collaborations, arXiv:1403.4427.
\bibitem{r18}CERN CMS,
 	CMS Collaboration,
 	CMS Collaboration
CMS-PAS-EXO-14-010, 	``Search for massive WH resonances decaying to ℓνbb¯ final state in the boosted regime at s√=8\,TeV''


\bibitem{1Bez} F.L.~Bezrukov, M.~Shaposhnikov, Phys. Lett. B{\bf 659}, 703 (2008).
\bibitem{2Bez} F.L.~Bezrukov, A.~Magnin, M.~Shaposhnikov, Phys. Lett. B{\bf 675}, 88
(2009).
\bibitem{3Bez} F.L.~Bezrukov and D.~Gorbunov, JHEP {\bf 1307}, 140 (2013),
arXiv:1303.4395.
\bibitem{4Bez} F.L.~Bezrukov and M.~Shaposhnikov,
J. Exp. Theor. Phys. {\bf 120}, 335 (2015) [Zh. Eksp. Teor. Fiz.
{\bf 147}, 389 (2015)], arXiv:1411.1923.
\bibitem{5Bez} F.~Bezrukov, J.~Rubio and M.~Shaposhnikov, {\it Living beyond the
edge: Higgs inflation and vacuum metastability}, Phys. Rev. D{\bf
92}8, 083512 (2015), arXiv:1412.3811.
\bibitem{1NLO}
 L.N.~Mihaila, J.~Salomon and M.~Steinhauser, Phys. Rev. Lett.
{\bf 108} (2012) 151602.
\bibitem{2NLO} K.~Chetyrkin and M.~Zoller, JHEP {\bf 06}, 033 (2012).
\bibitem{6Bez}
F.~Bezrukov, M.Yu. ~Kalmykov, B.A.~Kniehl, M.~Shaposhnikov, JHEP
{\bf 1210}, 140 (2012).
\bibitem{DIP}
C.D.~Froggatt, C.R.~Das, L.V.~Laperashvili, H.B.~Nielsen, Int. J.
Mod. Phys. A{\bf 30}, 21, 1550132 (2015); arXiv:1501.00139.
\bibitem{Bag1}
A.L. Macpherson and B.A. Campbell, Phys.Lett. B{\bf 306}, 379
(1993); ArXiv: hep-ph/9302278.
\bibitem{Bag2}
A.~Chodos, R.L.~Jaffe, K.~Johnson, C.B. ~Thorn and V.F.~Weisskopf,
Phys.Rev. D{\bf 9}, 3471 (1974).
\bibitem{Bag3}
W.A.~Bardeen, M.S.~Chanowitz, S.D.~Drell, M.~Weinstein and T.-M.
Yan, Phys.Rev. D{\bf 11}, 1094 (1975).
\bibitem{Kuch1} M.Yu.~Kuchiev, V.V.~Flambaum, E.~Shuryak, Phys. Rev. D{\bf 78},
077502 (2008); arXiv:0808.3632.
\bibitem{Kuch2} M.Yu.~Kuchiev, V.V.~Flambaum, E.~Shuryak, Phys. Lett. B{\bf
693}, 485 (2010); arXiv:0811.1387.
\bibitem{Kuch3} M.Yu.~Kuchiev, Phys. Rev. D{\bf 82}, 127701 (2010);
 arXiv:1009.2012.
\bibitem{Rich}
Jean-Marc Richard, {\it About the stability of the dodecatoplet},
Few Body Syst. {\bf 45}, 65 (2009); arXiv:0811.2711.

\bibitem{BS} H. Bethe, E. Salpeter (1951). "A Relativistic Equation for Bound-State 
Problems". Physical Review 84 (6): 1232. Bibcode:1951PhRv...84.1232S. doi:10.1103/PhysRev.84.1232.

\end{thebibliography}
\end{document}